# A Mobile Cloud-Based eHealth Scheme

Yihe Liu [1], Aaqif Afzaal Abbasi [2], Atefeh Aghaei [3], Almas Abbasi [4], Amir Mosavi [5,6,7], Shahab Shamshirband [8, 9, *] and Mohammed A. A. Al-qaness[10]

**Abstract:** Mobile cloud computing is an emerging field that is gaining popularity across borders at a rapid pace. Similarly, the field of health informatics is also considered as an extremely important field. This work observes the collaboration between these two fields to solve the traditional problem of extracting Electrocardiogram signals from trace reports and then performing analysis. The developed system has two front ends, the first dedicated for the user to perform the photographing of the trace report. Once the photographing is complete, mobile computing is used to extract the signal. Once the signal is extracted, it is uploaded into the server and further analysis is performed on the signal in the cloud. Once this is done, the second interface, intended for the use of the physician, can download and view the trace from the cloud. The data is securely held using a password-based authentication method. The system presented here is one of the first attempts at delivering the total solution, and after further upgrades, it will be possible to deploy the system in a commercial setting.

## 1 Introduction

The field of Mobile Cloud Computing (MCC) has been in the spotlight in recent times, primarily due to the popularity of mobile devices throughout the world. MCC can be considered as a hybrid between multiple technologies, and therefore, the field itself is quite complicated. The concept of cloud computing is relatively old and in the present, many companies (Google, Yahoo!, etc.) throughout the world are involved in the business of providing services via cloud computing. These services span from electronic mail, collaborative document sharing to health-care monitoring.

[1] Data Recovery Key Laboratory of Sichuan Province, College of Computer Science and AI, Neijiang Normal University, Neijiang, 641100, China.

[2] Department of Software Engineering, Foundation University, Islamabad, 44000, Pakistan.

[3] Department of Communication, Faculty of Social Sciences, University of Tehran, Tehran, 1417466191, Iran.

[4] Department of Computer Science, International Islamic University, Islamabad, 46000, Pakistan.

[5] Institute of Research and Development, Duy Tan University, Da Nang, 550000, Vietnam.

[6] Kalman Kando Faculty of Electrical Engineering, Obuda University, Budapest, 1034, Hungary.

[7] School of the Built Environment, Oxford Brookes University, Oxford, OX3 0BP, UK.

[8] Department for Management of Science and Technology Development, Ton Duc Thang University, Ho Chi Minh City, Vietnam.

[9] Faculty of Information Technology, Ton Duc Thang University, Ho Chi Minh City, Vietnam.

[10] School of Computer Science, Wuhan University, Wuhan, 430072, China.

* Corresponding Author: Shahaboddin Shamshirband. Email: shahaboddin.shamshirband@tdtu.edu.vn.

The field of Mobile Computing, similarly, is relatively old, however, the transformations that took place during the last decade in the field needs to be specially considered. These transformations can be attributed to the work of two key players, namely, Apple Inc. and Google Inc. With the invention of the iPhone (2007) that introduced the IOS to the world and Android OS (2008), the field of MCC has undergone a paradigm shift. Besides the two main operating systems, there are several other operating systems that are sharing a minor quota of the world smartphone users. The field of MCC also requires the collaboration of other fields such as wireless networking, data compression, security and privacy protocols [Abbasi, Abbasi, Shamshirband et al. (2019); Abbasi and Jin (2018)].

The aforementioned paradigm shift in the field of MCC has made it into one of the most valuable tools available to date. The potential contained within MCC is invaluable and surpasses the capabilities of many existing technologies. The bottleneck posed by the limited capabilities of mobile computing is completely eliminated by the use of the cloud system and due to this reason; it is possible to perform extremely complex calculations using smart-phones. Furthermore, the processing capabilities of smartphones themselves have evolved beyond expectation and today, almost all the computations that were once only possible using desktop computers can be performed by hand-held devices.

One major problem with using the cloud for computation purposes is the high data requirement and the associated charges by service providers. Customers throughout the world are reluctant towards using data facilities and would rather avoid it if possible. The aforementioned enhancements in the computational capabilities have allowed programmers and designers to almost completely eliminate this requirement. Storage of data, on the other hand, can be performed later, and due to the fact that modern devices have internet capabilities using Wi-Fi technology, the data cost problem is eliminated [Wang, Zhao and Hou (2018)].

Health informatics has been another field in the spotlight in the recent times, primarily due to the fact that it is one of the 14 grand challenges of engineering as proposed by the National Academy of Engineering [Shen, Zhou, Chen et al. (2017)]. Furthermore, the field itself is of vital importance to the future of the human race and therefore, is a field that is heavily invested all over the world. With the potential of MCC technology, it is obvious that the field of health informatics can gain considerable advantage by collaborating with it. This work attempts to address a traditional problem associated with health-care informatics with the idea of presenting a user-friendly platform to users for extraction and analysis of the electrocardiogram (ECG) signal.

The paper is organized as follows. Section 2 describes problems concerning cardiovascular diseases and electrocardiograms; Section 3 discusses the architecture and working of the presented framework. Section 4 presents evaluation and analysis of results. Finally, Section 5 concludes the paper.

## 2 Problem statement

### 2.1 Cardiovascular diseases and electrocardiograms

To understand the importance of extracting and analyzing ECG signals, it is important to analyze the current situation of cardiovascular diseases (CVDs) on a global scale. CVDs are popular as the leading cause of death throughout the world and for almost a century, the diagnoses of CVDs have relied upon the analysis of ECG records. While there have been several attempts at developing ECG machines that are portable and suitable for home use, the medical community is reluctant towards relying on the recordings made by these machines [Poon, Okin and Kligfield (2005)]. The other key parameter is that most of the machines in existence today are analog machines that are incapable of delivering a digitized record of the ECG signal. The report known as an ECG trace is printed on a graph paper and handed over to the patient for safekeeping. Due to the nature of these reports, it is not possible to keep a backup without additional effort and in clinical practice, it is often observed that patients tend to misplace these trace reports. Due to the chronic nature of CVDs, patients, as they age, are required to obtain an increasing number of ECG recordings and this often increases the severity of this issue. Further to the problem, the analog trace reports tend to show signs of fading away with time, and due to this reason, after a while the trace, even if it is preserved, becomes useless and unreadable [Xiong, Chen, Chen et al. (2019); Rupali and Karandikar (2018)].

### 2.2 Solution analysis

Upon observation, it is clear that the two solutions exist to this problem. The first solution is the replacement of analog scanners with digital scanners that are capable of giving the output in the form of a computer file that can be read by the physician using his computer. However, this is obviously not an easy solution and it would take decades before this solution is implemented. The second solution is presented through this paper where the user is given the front end application to photograph the ECG trace and the mobile computing component allows the user to extract the trace from the paper. Once the numerical signal is available to the user, it is transmitted to the cloud for further signal processing and storage purposes. A physician can be given access to cloud storage via mobile and desktop access for further analysis and diagnostic purposes. The signal extraction process greatly reduces the data transmission requirement and the online backup feature is extremely useful to the user [Rupali and Karandikar (2018)]. The rest of this paper describes the methodology used to extract, analyze and store the electrocardiogram signal.

## 3 Methodology

### 3.1 Photographing the ECG trace

Most modern smart-phones are equipped with a camera that can be used to photograph the ECG trace upon generation. A sensitive signal of this sort needs to be correctly extracted in order to avoid misdiagnoses and other potential problems. Due to this reason, it is vital that the photograph is obtained from an acceptable camera. For experimentation purposes, the Moto G (1st generation) mobile device by Motorola Inc. running an Android 4.4.4 is used. It has a 5 MP camera which is perfect to capture low-density targets for trace extraction

[Barbera, Kosta, Mei et al. (2013)].

All photographs were taken in good lighting conditions and the traces were placed as horizontally as possible. Once the photograph is taken, a simple algorithm based on the Radon transform is used to straighten the image [Kar, Saha, Bera et al. (2019)]. This process is almost instantaneous and the required algorithms are available via the OpenCV library for Android. Once the image is straightened, the user is given the option to crop the image in case the background appears to be distracting the extraction process. In practice, multicolored backgrounds often tend to leave artifacts on the image in the next set of steps.

## 3.2 Extracting the ECG signal

It is necessary to convert the color image into the gray-scale as the first step. For this task, the standard ITU algorithm shown in Eq. (1) is applied to the image [Yu and Jung (2019)]. The terms R, G, and B correspond to the three layers of the 8-bit color image and Y corresponds to the gray-scale layer.

$$Y_i = 0.299 \times R_i + 0.587 \times G_i + 0.114 \times B_i, (\forall i \in I) \tag{1}$$

Afterward, the image is converted into the morphology domain using Otsu's method. The Otsu method is a histogram-based method that determines a threshold based on the separation of the grey-levels of the image. This dynamic threshold is one of the most popular methods used in image analysis to convert a gray-scale image into a morphological image [Bengio, Courville and Vincent (2013)]. All these methods are available via the OpenCV library and therefore, can directly be applied to the photograph without tedious coding.

Once the morphological image is available, it is essential to remove unnecessary artifacts using artifact removal techniques. Using the 2-D convolution method, the image is convoluted with a 4-pixel long linear structuring element across the entire angular span from 0 to 180 degrees. The noise removal binary trace can then be used to detect the signal [Anitescu, Atroshchenko, Alajlan et al. (2019)].

Signal extraction is best attempted by using an envelope detection method where the image is scanned across each column. This extraction is shown in Fig. 1. At each column, binary positives are searched from both the top and the bottom of the column, and the first appearance of a binary positive in each direction is extracted as belonging to the top and bottom envelopes respectively. In instances where a pixel is missing, the value can be either interpolated from the previous and next values or the same value from the previous column can be repeated. Based on empirical results, both methods yield similar results. Once the two envelopes are extracted, the average of the two envelopes can be considered as the signal value. The simple envelope detection algorithm is the only portion of this system that needs to be coded in the front end and once this is complete, the extraction task is complete [Rupali and Karandikar (2018); Khanna, Iyer and Vetter (2019); Sorokina, Sergeyev and Turitsyn (2019)]

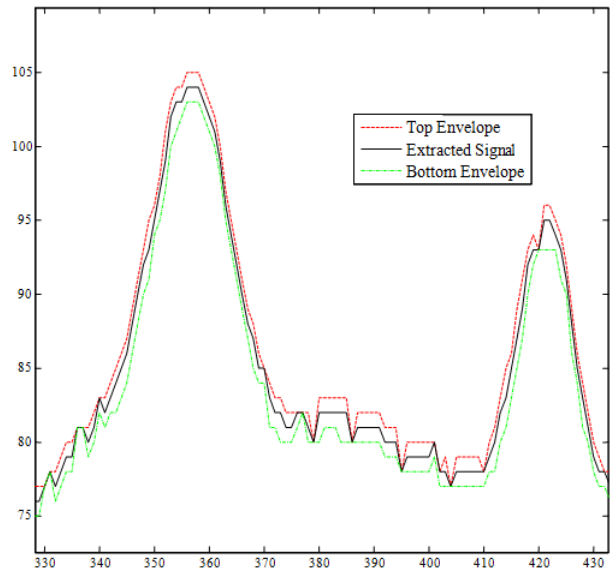

**Figure 1:** Envelope detection

*3.3 Mapping from image domain to signal domain*

The extracted signal is valued on the basis of the geometric position of the pixels and needs to be converted back to the signal domain. Mathematically, this conversion depends upon multiple variables such as the number of megapixels of the camera sensor and the distance from the trace to the camera, etc.

However, it is not possible to restrict these parameters in general use. Therefore, as an alternative, the width of the trace is used for calibration purposes. In the front-end itself, the user is requested to decide the height of the signal trace area. Since each trace paper is of standard height (3.5 cm), it is possible to calculate the distance represented by each pixel value. By multiplying the geometric height of the pixel with this computed coefficient, it is possible to backtrack the original value of the signal for testing results success.

*3.4 Post processing tasks*

Once the signal is successfully extracted, it is shown to the users for verification. Once the accuracy is verified, the signal trace is backed up in the memory (photograph) as well as in terms of signal values. Once the mobile device establishes internet access, the signal values are uploaded onto a local server where further processing tasks are carried out. These back-end tasks include the removal of signal noise (if present) using multiple filters such as direct FIR filters and Savitzky-Golay filters [Khorasani, Hodtani and Kakhki (2019)]. Furthermore, algorithms such as Pan-Tompkins can also be used to detect the R-peaks of the signal trace [Tang, Hu and Tang (2018)]. In addition to the above, processing includes the detection of the heart rate, the standard deviation of R-peaks as well as the detection of other components such as P, Q, S and T waves. These values, although not implemented in this iteration, can be used to make predictions about the ECG and if necessary, warn users about potential abnormalities that might need attention [Rupali and Karandikar (2018), Khanna, Iyer and Vetter (2019); Sana, Ballal, Shadaydeh et al. (2019); Roland, Wimberger, Amsuess et al. (2019)]. The back-end also provides an additional service with the support of different front-end software that is provided to the physician.

This application can log in to the server and based on authentication parameters provided to the user, the user is given the ability to allow the physician to access ECG signals from its cloud platform. Since these signal trace files are extremely small (ranging only a few kilobytes), even mobile data connection is sufficient for performing operations. Once downloaded, the physician can analyze and view the files through his own device and make further analysis and decision making regarding the patient. The completed system is shown in Fig. 2 is a block diagram and Fig. 3 shows the extracted and analyzed signal for a lead 2 trace. The entire back-end at the server is supported by the MATLAB platform for the analysis part.

**4 Experiments and results**

In this section, we first describe the methodology used to test the performance of the proposed scheme. Since the proposed method might contain anomalies (often known as signal noise), we needed to verify the accuracy of various features points. In order to clearly identify the traces, a manual system can also be adopted. However, it is very laborious. Another method to perform the testing is to develop a generality area of existence on graph note to clearly identify the area of feature point presence. We use the following methods for precision determination Eq. (2) and precision add up recall Eq. (3) in the proposed scheme for fine tuning the feature point delineation and abnormality.

$$\text{Precision value} = \frac{\{\text{Total feature points}\} \cap \{\text{Selected feature points}\}}{\{\text{Total selected feature points}\}} \qquad (2)$$

$$\text{Precision recall} = \frac{\{\text{Selected feature points}\} \cap \{\text{Total feature points}\}}{\{\text{Selected feature points}\}} \qquad (3)$$

A collection of 50 traces photographed from the aforementioned device were used in the experiment. The photographs were captured in slightly varying lighting conditions and 5 traces were deliberately captured with geometric distortions where the camera was not held parallel to the object. The extraction was successful in 32 of the traces without any intervention from the user. The remaining 18 traces had visible problems amounting to discrepancies between the photograph and the extracted signal. The 5 traces with geometric distortions as expected were extracted erroneously.

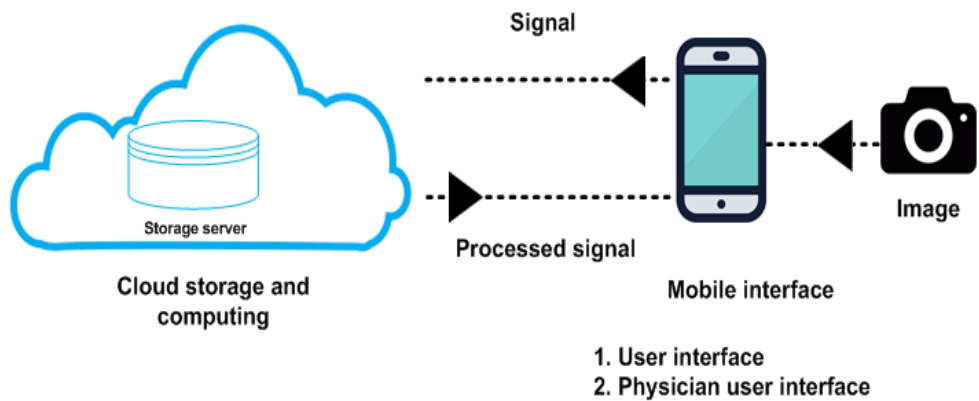

**Figure 2:** System overview

Regarding the 32 traces that were successfully extracted using the above method, only 18 belonged to the lead II and complete analysis was performed on other 18 traces. Algorithms used in the back-end delivered almost 99 percent accuracy results as expected with only a few missing R-peaks. Since the detection of all the other parameters depends on the correct detection of all the R-peaks, 2 traces were discarded at this stage. The detection of other peaks and waves (P, Q, S, and T) were successful for 94.3 percent of the peaks in the 16 traces. Since the heart rate, as well as the standard deviation values, only depends on the extraction of the R-peak, these parameters were not affected by this inaccuracy.

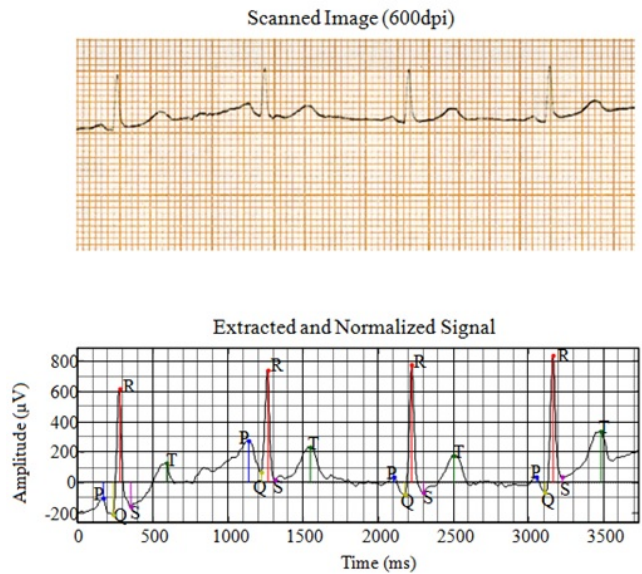

**Figure 3:** Photograph and analyzed signal trace

**5 Conclusion**

This research is one of the first attempts at extracting and analyzing the ECG trace using a mobile system. Although many attempts have been made at extracting the ECG trace using desktop computers and standard scanners, the task at hand using a smart-phone camera is much more challenging. The scanner always scans at a known resolution and the distortion of images is never an issue. In the case of a hand-held camera, many unknown parameters play a role.

This research has not addressed all the various image processing parameters that need to be carefully analyzed to successfully perform the extraction task. In most of the cases, the reason for failure is the lower success rates at extraction. Furthermore, it is best to use a toolkit approach at extracting signals from traces of this nature due to the high trace variability. The performed research proves that mobile computing is a valid and flexible approach to tackling ECG extraction and analysis problems. By developing these algorithms, these applications can be commercialized in the near future.

**Acknowledgment:** We thank anonymous reviewers for their feedback which helped in the improvement and presentation of this article. We also acknowledge Professor Rajkumar Buyya (The University of Melbourne, Australia) for his feedback on an earlier version of the manuscript.

**Conflicts of Interest:** The authors declare that they have no conflicts of interest to report regarding the present study.